\documentclass[conference]{IEEEtran}
\IEEEoverridecommandlockouts
\usepackage{amsfonts}
\usepackage{amsmath}
\usepackage{algorithmic}
\usepackage[boxed]{algorithm}
\usepackage{mathrsfs}
\usepackage{graphicx}

\begin{document}
\title{Joint Transmitter-Receiver Design for the Downlink Multiuser Spatial Multiplexing MIMO System}

\author{\IEEEauthorblockN{Pengfei~Ma, Wenbo~Wang, Xiaochuan~Zhao and Kan~Zheng}
\IEEEauthorblockA{Wireless Signal Processing and Network Lab, \\
Key Laboratory of Universal Wireless Communication Ministry of
Education, \\
Beijing University of Posts and Telecommunications,
Beijing, China}}

\maketitle

\begin{abstract}
%\boldmath
In the multiuser spatial multiplexing multiple-input multiple-output
(MIMO) system, the joint transmitter-receiver (Tx-Rx) design is
investigated to minimize the weighted sum power under the
post-processing signal-to-interference-and-noise ratio (post-SINR)
constraints for all subchannels. Firstly, we show that the
uplink-downlink duality is equivalent to the Lagrangian duality in
the optimization problems. Then, an iterative algorithm for the
joint Tx-Rx design is proposed according to the above result.
Simulation results show that the algorithm can not only satisfy the
post-SINR constraints, but also easily adjust the power distribution
among the users by changing the weights accordingly. So that the
transmitting power to the edge users in a cell can be decreased
effectively to alleviate the adjacent cell interference without
performance penalty.
\end{abstract}


\IEEEpeerreviewmaketitle

\section{Introduction}
Spatial multiplexing for the multiple-input multiple-output (MIMO)
systems, employing multiple transmit and receive antennas, has been
recognized as an effective way to improve the spectral efficiency of
the wireless link \cite{1}. More recently, the multiuser schemes
have been investigated for the spatial multiplexing MIMO systems.
This paper focuses on the downlink multiuser schemes in which each
user can not cooperate with the others thus suffers from the
interference from them.

Mainly, there are two kinds of multiuser schemes. One is the
precoder or the transmit beamforming, such as the dirty-paper coding
(DPC) \cite{2} and the zero-forcing (ZF) \cite{3}, etc., which
mitigates the multiuser interference only by processing at the
transmitter. The other is the joint transmitter-receiver (Tx-Rx)
design, such as the nullspace-directed SVD (Nu-SVD) \cite{4} and the
minimum total mean squared error (TMMSE) \cite{5}, etc. In general,
the former possesses lower complexity but more performance penalty.
With the great development of signal processors, the latter
gradually draws more attention.

For the joint Tx-Rx design, the schemes proposed in \cite{4}\cite{5}
minimize mean squared error (MMSE), or maximize the capacity under
the transmit power constraint. Whereas on some occasions, such as
the multimedia communication, it is required to minimize the total
transmit power while guarantee the quality of service (QoS).
\cite{6}\cite{7} investigate the beamforming and the power
allocation policy when all users are subjected to a set of
post-processing signal-to-interference-and-noise ratio (post-SINR)
constrains in the uplink SIMO and the downlink MISO.
\cite{8}\cite{9} extend this work to the downlink MIMO and the MIMO
network, however the MIMO systems discussed in \cite{8}\cite{9} are
assumed that there is only one substream between each pair of the
transmitter and receiver. In other words, only the multiuser
interference appears in the so-called diversity MIMO system in
\cite{8}\cite{9}. For the multiuser spatial multiplexing MIMO
system, however, both the multiuser interference between individual
users and self-interference between individual substreams of a user
should be mitigated.

For the downlink, the transmit beamforming affects the interference
signature of all receivers, whereas the receive beamforming only
affects that of the corresponding user. \cite{7}\cite{8} construct a
dual system, called the virtual uplink, and indicate that the
virtual uplink can obtain the same post-SINR as the primary
downlink. Moreover, the receive beamforming matrix of the virtual
uplink is identical with the transmit beamforming matrix of the
primary downlink. The design of the downlink, therefore, can resort
to the virtual uplink.

In this paper, we extend the duality derived for MIMO network in
\cite{9} to the multiuser spatial multiplexing MIMO system.
According to the uplink-downlink duality, we propose a joint Tx-Rx
scheme to minimize the weighted sum power under the post-SINR
constraints of all the subchannels.

\emph{Notation}: Boldface upper-case letters denote matrices, and
boldface lower-case letters denote column vectors. $tr( \cdot )$, $(
\cdot )^*$, $( \cdot )^H$, $|| \cdot ||_2$ and $|| \cdot ||_F$
denote trace, conjugate, conjugate transposition, Euclidian norm and
Frobenius norm, respectively. $diag({\bf{x}})$ denotes a diagonal
matrix with diagonal elements drawn from the vector ${\bf{x}}$.
$[{\bf{ \cdot }}]_{i,j}$, $[{\bf{ \cdot }}]_{{\bf{:}},j}$ denote the
$(i$,$j)$-th element and $j$-th column of a matrix, respectively.


\section{System model}
We consider a base station (BS) with $M$ antennas and $K$ mobile
stations (MS's) each having $N_i(i=1,\ldots,K)$ antennas. There are
$L_i(i=1,\ldots,K)$ substreams between BS and MS$_i(i=1,\ldots,K)$,
that is to say, BS transmits $L_i$ symbols to MS$_i$ simultaneously.
The signal recovered by MS$_k$ can be written as
\begin{equation}
 {\bf{y}}_k ^{DL}  = {\bf{A}}_k ^H {\bf{H}}_k
\sum\limits_{i = 1}^K {{\bf{B}}_i diag(\sqrt {{\bf{p}}_i }
){\bf{x}}_i }  + {\bf{A}}_k ^H {\bf{n}}_k
\end{equation}
where $\mathbf{y}_{k}^{DL}\in \mathcal{C}^{L_k\times1}$ is the
recovered signal vector. $\mathbf{x}_{i}\in
\mathcal{C}^{L_{i}\times1}(i=1,\ldots,K)$ is the transmitted signal
vector from BS to MS$_{i}$ with zero-mean and normalized covariance
matrix $\mathbf{I}$. $\mathbf{p}_{i}\in\mathcal{R}^{L_{i}\times1}$
denotes the power vector allocated to MS$_{i}$. A linear post-filter
$\mathbf{A}_{k}\in\mathcal{C}^{N_k\times{L_k}}$ is used to recover
an estimation of the transmitted signal vector $\mathbf{x}_{k}$. The
MIMO channel from BS to MS$_{k}$ is denoted as $\mathbf{H}_{k}\in
\mathcal{C}^{N_{k}\times M}$, and assumed flat faded. Hence, its
elements are the complex channel gains, and they are independently
identically distributed (i.i.d.) zero-mean complex Gaussian random
variables with the unity variance. Moreover, the perfect channel
state information are assumed available at both transmitter and
receiver via some way, for example, channel measurement at receiver
and fast feedback to the transmitter for the frequency division
duplex (FDD) systems, or invoking the channel reciprocity in time
division duplex (TDD) systems. $\mathbf{B}_{i}\in \mathcal{C}^{M
\times L_{i}}$ is used to weight $\mathbf{x}_{i}$ and transform it
into a $M\times1$ vector. $\mathbf{n}_k\in \mathcal{C}^{N_k \times
1}$ is the noise vector with the correlation matrix
$\mathbf{R}_{n}=\sigma_n^{2}\mathbf{I}$. For simplicity, in the
sequel we assume $L_{1}=\ldots=L_{K}=L$.

We design the $\mathbf{A}_{k}$, $\mathbf{B}_{k}$ and $\mathbf{p}_{k}
(k=1,\ldots,K)$ in (1) to minimize the weighted sum power under the
post-SINR constraints, which can be denoted as the following
optimization problem.
\begin{equation}
\begin{array}{cl}
   {\mathop {\min }\limits_{{\bf{p}},{\bf{A}}_k ,{\bf{B}}_k } } & {{\bf{w}}^T {\bf{p}}} \\
   {s.t.} & {SINR_{k,j} ^{DL}  \ge \gamma _{k,j} }
\end{array}
 {(k=1,\ldots,K, j=1,\ldots,L)}
\end{equation}
where ${\bf{p}} = [{\bf{p}}_1^T,\ldots,{\bf{p}}_K^T]^T$ and
$\mathbf{w}\in\mathcal{R}^{KL\times1}$ is the weight vector.
$\mathbf{w}$ affects the power distribution among users, and its
value is determined by various factors, such as the positions of
users in a cell and the interference environment of the neighboring
cells.  $\gamma _{k,j}$
 is the given post-SINR goal for the MS$_{k}$'s $j$-th substream.

\section{The proof of uplink-downlink duality}
If
${\bf{A}}_k=\left[{{\bf{a}}_{k,1}},{\ldots},{{\bf{a}}_{k,L}}\right]$,
${\bf{B}}_k=\left[{{\bf{b}}_{k,1}},{\ldots},{{\bf{b}}_{k,L}}\right]$,
${\bf{p}}_k=\left[{p_{k,1}},{\ldots},{p_{k,L}}\right]^T$, (1) can be
rewritten into
\begin{equation}
\begin{aligned}
\!\!\!\!\!\!{\bf{y}}_k^{DL}&=\!\!\left[\!{\begin{array}{*{4}c}
   {{\bf{a}}_{k,1}^H{\bf{H}}_k{\bf{b}}_{k,1}\sqrt{p_{k,1}}}&\!\!{\ldots}&\!\!{{\bf{a}}_{k,1}^H{\bf{H}}_k{\bf{b}}_{k,L}\sqrt{p_{k,L}}}\\
   {\vdots}&\!\!{\ddots}&\!\!{\vdots}\\
   {{\bf{a}}_{k,L}^H{\bf{H}}_k{\bf{b}}_{k,1}\sqrt{p_{k,1}}}&\!\!{\ldots}&\!\!{{\bf{a}}_{k,L}^H{\bf{H}}_k{\bf{b}}_{k,L}\sqrt{p_{k,L}}}\\
\end{array}}\!\right]{\bf{x}}_k\\
   \!\!\!\!\!\!{}&+{\bf{A}}_k^H{\bf{H}}_k\sum\limits_{i=1,i{\ne}k}^K{{\bf{B}}_{i}diag(\sqrt{{\bf{p}}_i}){\bf{x}}_i}+{\bf{A}}_k^H{\bf{n}}_k
\end{aligned}
\end{equation}
The diagonal elements of the first part in the right-hand side (RHS)
of (3) denote the useful signals, and the non-diagonal elements
denote the self-interference. The medial and the last parts in the
RHS of (3) denote the multiuser interference and the noise,
respectively. Moreover, the post-SINR of the MS$_{k}$'s $j$-th
substream can be denote as
\begin{equation}
\begin{aligned}
 SINR_{k,j} ^{DL} & = \frac{{{\bf{a}}_{k,j} ^H {\bf{R}}_{k,j}^{s,DL} {\bf{a}}_{k,j} }}{{{\bf{a}}_{k,j} ^H {\bf{R}}_{k,j}^{I + n,DL} {\bf{a}}_{k,j} }} \\
 {\bf{R}}_{k,j}^{s,DL} & = p_{k,j} {\bf{H}}_k {\bf{b}}_{k,j} {\bf{b}}_{k,j} ^H {\bf{H}}_k ^H  \\
 {\bf{R}}_{k,j}^{I + n,DL} & = \sum\limits_{i = 1,i \ne j}^L {p_{k,i} {\bf{H}}_k {\bf{b}}_{k,i} {\bf{b}}_{k,i} ^H } {\bf{H}}_k ^H  +  \\
     & \sum\limits_{m = 1,m \ne k}^K {{\bf{H}}_k {\bf{B}}_m diag({\bf{p}}_m ){\bf{B}}_m ^H } {\bf{H}}_k ^H  + \sigma _n ^2
     {\bf{I}} \\
\end{aligned}
\end{equation}
If ${\bf{x}}_m=\left[{x_{m,1} },{\ldots},{x_{m,L}}\right]$,
${\bf{y}}_k ^{DL}=\left[{y_{k,1}},{\ldots},{y_{k,L}}\right]$, the
link power gain between $x_{m,n}$ and $y_{k,j}$ can be denoted as
\begin{equation}
[{\bf{\phi }}_{k,j} ]_{m,n}  = ||{\bf{a}}_{k,j} ^H {\bf{H}}_k
{\bf{b}}_{m,n} ||_2^2
\end{equation}
then (4) can be rewritten into
\begin{equation}
\begin{array}{l}
 SINR_{k,j} ^{DL}  =  \\
 \frac{{p_{k,j} [{\bf{\phi }}_{k,j} ]_{k,j} }}{{\sum\limits_{i = 1,i \ne j}^L {p_{k,i} [{\bf{\phi }}_{k,j} ]_{k,i} }  + \sum\limits_{m = 1,m \ne k}^K {\sum\limits_{n = 1}^L {p_{m,n} [{\bf{\phi }}_{k,j} ]_{m,n} }  + \sigma _n ^2 ||{\bf{a}}_{k,j}||_2^2 } }} \\
 \end{array}
\end{equation}
By substituting (6) into the constraint inequality of (2), we obtain
\begin{equation}
{\bf{c}}_{k,j} ^T {\bf{p}} + \sigma _n ^2 ||{\bf{a}}_{k,j}||_2^2 \le 0 \;\;\; {(k = 1,\ldots,K,j = 1,\ldots,L)}\\
\end{equation}
where the $m$-th element of
$\mathbf{c}_{k,j}\in\mathcal{R}^{KL\times1}$ is
\begin{equation}
\left[{{\bf{c}}_{k,j}}\right]_m=\left\{
\begin{array}{cc}
-\frac{{[{\bf{\phi}}_{k,j}]_{k,j}}}{{\gamma_{k,j}}} & m=(k-1)L+j\\
{[{\bf{\phi}}_{k,j}]}_{\left\lceil{\frac{m}{L}}\right\rceil,m-(\left\lceil{\frac{m}{L}}\right\rceil-1)L}&m\ne(k-1)L+j\\
\end{array}\right.
\end{equation}
where $\left\lceil {\frac{m}{L}} \right\rceil$ rounds $\frac{m}{L}$
to the nearest integer greater than or equal to $ \frac{m}{L}$ .
Write (7) into the matrix form, we obtain
 \begin{equation}
{\bf{Cp}} + {\bf{d}} \le 0
\end{equation}
where $\mathbf{C}\in\mathcal{R}^{KL\times KL}$ and
$\mathbf{d}\in\mathcal{R}^{KL\times 1}$ are
\begin{equation}
\begin{array}{lll}
{\bf{C}}&=&\left[{{\bf{c}}_{1,1}},{\ldots},{{\bf{c}}_{1,L}},{\ldots},{{\bf{c}}_{K,1}},{\ldots},{{\bf{c}}_{K,L}}\right]^T \\
{\bf{d}}&=&\sigma_n^2\left[{||{\bf{a}}_{1,1}||_2^2},..,{||{\bf{a}}_{1,L}||_2^2},..,{||{\bf{a}}_{K,1}||_2^2},..,{||{\bf{a}}_{K,L}||_2^2}\right]^T\\
\end{array}
\end{equation}
So, (2) is equivalent to the following optimization problem
 \begin{equation}
\begin{array}{cl}
   {\mathop {\min }\limits_{{\bf{p}}_k ,{\bf{A}}_k ,{\bf{B}}_k } } & {{\bf{w}}^T {\bf{p}}}  \\
   {s.t.} & {{\bf{Cp}} + {\bf{d}} \le 0}\, , \,\,\, {{\bf{p}} \ge 0}  \\
 \end{array}
\end{equation}
Subsequently, to obtain the Lagrangian duality of (11) \cite{9}, we
divide the solving process of (11) into two steps similar with
\cite{10}. First, assuming ${\bf{A}}_k $ and ${\bf{B}}_k(k =
1,\ldots,K)$ are fixed, the Lagrangian function of (11) is
 \begin{equation}
L({\bf{p}},{\mbox{\boldmath{$\lambda$}}},{\mbox{\boldmath{$\mu$}}})={\bf{w}}^T
{\bf{p}} + {\mbox{\boldmath{$\lambda$}}}^T ({\bf{Cp}} + {\bf{d}}) -
{\mbox{\boldmath{$\mu$}}}^T {\bf{p}}
\end{equation}
where ${\mbox{\boldmath{$\lambda$}}} \ge {\bf{0}}$,
${\mbox{\boldmath{$\mu$}}} \ge {\bf{0}}$ are the Lagrangian
multipliers associated with the inequality constraints. Then the
Lagrangian duality of (11) is
\begin{equation}
\begin{array}{cl}
 {\mathop {\max }\limits_{{\boldmath{\lambda}},{\boldmath{\mu}}} \mathop {\min }\limits_{\bf{p}}} & L({\bf{p}},{\mbox{\boldmath{$\lambda$}}},{\mbox{\boldmath{$\mu$}}}) \\
 {s.t.} & {{\mbox{\boldmath{$\lambda$}}} \ge {\bf{0}}}\; ,\; {{\mbox{\boldmath{$\mu$}}} \ge {\bf{0}}}  \\
\end{array}
\end{equation}
According to the Slater's condition, (11) is equivalent to (13).
Since the gradient of the Lagrangian function (12) with respect to
$\bf{p}$ vanishs at optimal points, we obtain ${\bf{w}}^T  -
{\mbox{\boldmath{$\mu$}}}^T =  - {\mbox{\boldmath{$\lambda$}}}^T
{\bf{C}}$. Substituting it into (12), we obtain $\mathop {\min
}\limits_{\bf{p}}
L({\bf{p}},{\mbox{\boldmath{$\lambda$}}},{\mbox{\boldmath{$\mu$}}})
= {\bf{d}}^T {\mbox{\boldmath{$\lambda$}}}$. Moreover, as
${\mbox{\boldmath{$\lambda$}}}\ge{\bf{0}}$ and
${\mbox{\boldmath{$\mu$}}}\ge{\bf{0}}$, (13) can be rewritten to
\begin{equation}
\begin{array}{cl}
   {\mathop {\max }\limits_{\bf{\lambda }} } & {{\bf{d}}^T {\mbox{\boldmath{$\lambda$}}}}  \\
   {s.t.} & {{\bf{C}}^T {\mbox{\boldmath{$\lambda$}}} + {\bf{w}} \ge {\bf{0}}}  \\
   {} &  {{\mbox{\boldmath{$\lambda$}}} \ge {\bf{0}}}  \\
\end{array} \\
\end{equation}
Similar with (6)-(9), substitute (10) into (14), we obtain
 \begin{equation}
\begin{array}{cl}
   {\mathop {\max }\limits_{\bf{\lambda }} } & {{\bf{d}}^T {\mbox{\boldmath{$\lambda$}}}}  \\
   {s.t.} & {SINR_{k,j} ^{UL}  \le \gamma _{k,j} }  \\
\end{array}
 {(k = 1,\ldots,K,j = 1,\ldots,L)}
\end{equation}
where
\begin{equation}
\begin{aligned}
SINR_{k,j} ^{UL} &= \frac{{{\bf{b}}_{k,j} ^H {\bf{R}}_{k,j}^{s,UL} {\bf{b}}_{k,j} }}{{{\bf{b}}_{k,j} ^H {\bf{R}}_{k,j}^{I + n,UL} {\bf{b}}_{k,j} }} \\
{\bf{R}}_{k,j}^{s,UL}  &= \lambda _{k,j} {\bf{H}}_k {\bf{a}}_{k,j} {\bf{a}}_{k,j} ^H {\bf{H}}_k ^H  \\
{\bf{R}}_{k,j}^{I + n,UL}  &= \sum\limits_{i = 1,i \ne j}^L {\lambda _{k,i} {\bf{H}}_k {\bf{a}}_{k,i} {\bf{a}}_{k,i} ^H {\bf{H}}_k ^H }  \\
& + \sum\limits_{m = 1,m \ne k}^K {{\bf{H}}_m {\bf{A}}_m diag({\mbox{\boldmath{$\lambda$}}}_m ){\bf{A}}_m ^H {\bf{H}}_m ^H } \\
& + [{\bf{w}}]_{(k-1)L+j}{\bf{I}} \\
\end{aligned}
\end{equation}
where
${\mbox{\boldmath{$\lambda$}}}=[{\mbox{\boldmath{$\lambda$}}}_1^T,\ldots,{\mbox{\boldmath{$\lambda$}}}_K^T]$.
Furthermore, $SINR_{k,j} ^{UL}$ is the post-SINR of MS$_{k}$'s
$j$-th substream in the virtual uplink
 \begin{equation}
{\bf{y}}_k ^{UL}  = {\bf{B}}_k ^H \sum\limits_{i = 1}^K {{\bf{H}}_i
{\bf{A}}_i diag(\sqrt {{\mbox{\boldmath{$\lambda$}}}_i } ){\bf{x}}_i
}  + {\bf{B}}_k^H \sqrt{{\bf{w}}_k }
\end{equation}
(14) maximizes the weighted sum power under the maximum post-SINR
constraints, however, it has no physical meaning \cite{9}. But it
can be shown that (14) is equivalent to the following optimization
problem
 \begin{equation}
\begin{array}{cl}
   {\mathop {\min }\limits_{\bf {\lambda}}} & {{\bf{d}}^T {\mbox{\boldmath{$\lambda$ }}}}  \\
   {s.t.} & {SINR_{k,j} ^{UL}  \ge \gamma _{k,j} }  \\
 \end{array}
 {(k = 1,\ldots,K,j = 1,\ldots,L)}
\end{equation}

\emph{Theorem 1}: At the optimal point, the post-SINR constraints in
(15) and (18) are active. And the solutions of (15) and (18) are
identical.

\emph{Proof}: Without any loss of the generality, we assume $
SINR_{k,j} ^{UL}  < \gamma _{k,j}$. From (16), we can find $\lambda
_{k,j}$ contribute to the numerator of $SINR_{k,j} ^{UL}$ and the
denominator of $SINR_{m,n} ^{UL} (m \ne k,n \ne j)$. In other words,
$SINR_{k,j} ^{UL}$ is a monotone increasing function of $\lambda
_{k,j}$, while $SINR_{m,n} ^{UL} (m \ne k,n \ne j)$ is a monotone
decreasing function of $\lambda _{k,j}$. So increasing $\lambda
_{k,j}$ until $SINR_{k,j} ^{UL}  = \gamma _{k,j}$,  we obtain a
larger ${\bf{d}}^T {\mbox{\boldmath{$\lambda$}}}$ without breaking
any post-SINR constraint. Likewise, if $SINR_{k,j} ^{UL}  > \gamma
_{k,j}$, decreasing $\lambda _{k,j}$ until $SINR_{k,j} ^{UL}  =
\gamma _{k,j}$, a smaller ${\bf{d}}^T {\mbox{\boldmath{$\lambda$}}}$
is obtained. As a result, the constraints of (15) and (18) become a
linear equations ${\bf{C}}^T {\mbox{\boldmath{$\lambda$}}} +
{\bf{w}} = {\bf{0}}$, and its solution is
${\mbox{\boldmath{$\lambda$}}}^*  =  - ({\bf{C}}^T )^{ - 1}
{\bf{w}}$.

Similar with \emph{Theorem 1}, at the optimal point of (11)
${\bf{p}}^*  = - {\bf{C}}^{ - 1} {\bf{d}}$.

Summarize the above statement, we obtain the following conclusion.

\emph{Theorem 2}: In the downlink multiuser spatial multiplexing
MIMO system, if the transmit and receive beamforming matrices are $
{\bf{B}}_k$ and ${\bf{A}}_k ^H (k = 1,\ldots,K)$, respectively, as
long as the following conditions are satisfied, the downlink
optimization problem (2) is equivalent to the virtual uplink
optimization problem (18).

1) In the virtual uplink, the transmit and receive beamforming
matrices are ${\bf{A}}_k$ and ${\bf{B}}_k ^H (k = 1,\ldots,K)$,
respectively.

2) In the virtual uplink problem (18), the weight vector ${\bf{w}}$
is the noise power vector.

3) In the virtual uplink problem (18), the noise power vector
${\bf{d}}$ is the weight vector.

When ${\bf{A}}_k$ and ${\bf{B}}_k (k = 1,\ldots,K)$ are not fixed,
(18) is a joint optimization problem denoted as
\begin{equation}
\begin{array}{cl}
   {\mathop {\min }\limits_{{\bf{\lambda }},{\bf{A}}_k ,{\bf{B}}_k } } & {{\bf{d}}^T {\mbox{\boldmath{$\lambda$}}}}  \\
   {s.t.} & {SINR_{k,j} ^{UL}  \ge \gamma _{k,j} }  \\
\end{array}
 {(k = 1,\ldots,K,j = 1,\ldots,L)}
\end{equation}

\emph{Theorem 3}: If the noise power vector in the virtual uplink is
the weight vector ${\bf{w}}$, the joint optimization problem (2) is
equivalent to (19). At the optimal point, the beamforming matrices
of the virtual uplink and the primal downlink are common.

\emph{Proof}:  Let
${({\bf{B}}_k^*,{\bf{A}}^*_k,{\bf{p}}^*)}\;(k=1,\ldots,K)$ be the
global minimum of (2). According to \emph{Theorem 2}, (19) has the
solution
${({\bf{A}}_k^*,{\bf{B}}^*_k,{\mbox{\boldmath{$\lambda$}}}^*)}\;{k=1,\ldots,K}$.
Moreover, this solution is definitely the global minimum. Otherwise,
a better solution of (2) would be found by applying \emph{Theorem 2}
again. So the virtual uplink and the primal downlink have the common
beamforming matrices.

The weight vector ${\bf{w}}$ decides whether ${\bf{A}}_k $ and
${\bf{B}}_k$ are used to strengthen the useful signals or alleviate
the interference to other users. When a user's weight turns higher,
its transmit power will decrease. In this occasion, it benefits to
apply the beamforming to increase the signal gain, as the
interference to other users is much less important. On the other
hand, once the weight gets lower, the beamformer should try to
suppress interference to others \cite{9}.

To mitigate the adjacent cell interference, we can increase the
weights of edge users in a cell, which would induce the declining of
the transmit power from the BS to them. In order to hold the
post-SINR under this circumstance, obviously, the beamforming
matrices would be used to boost up the signal gain.

\section{The joint Tx-Rx beamforming scheme}
It is rather difficult to solve the joint optimization problem (2)
directly. However, it is easy to obtain ${\bf{A}}_k (k =
1,\ldots,K)$ in the primal downlink, and so does ${\bf{B}}_k (k =
1,\ldots,K)$ in the virtual uplink. Moreover, it is proved in the
previous section that the primal downlink is equivalent to the
virtual uplink, and they have the common beamforming matrices
${\bf{A}}_k$ and ${\bf{B}}_k (k = 1,\ldots,K)$. Therefore, we divide
the solving process into four steps shown as Fig. \ref{Fig 1}.

\setlength{\unitlength}{1.5cm}
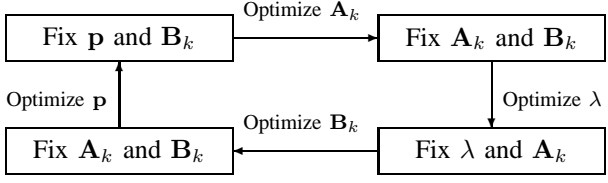
\begin{figure}
\begin{picture}(4,2)
\put(0.2,1){\framebox(2,0.4){Fix ${\bf{p}}$ and ${\bf{B}}_k$}}
\put(3.5,1){\framebox(2,0.4){Fix ${\bf{A}}_k$ and ${\bf{B}}_k$}}
\put(0.2,0){\framebox(2,0.4){Fix ${\bf{A}}_k$ and ${\bf{B}}_k$}}
\put(3.5,0){\framebox(2,0.4){Fix ${\bf{\lambda}}$ and ${\bf{A}}_k$}}
\put(2.2,1.2){\vector(1,0){1.3}} \put(4.5,1){\vector(0,-1){0.6}}
\put(3.5,0.2){\vector(-1,0){1.3}} \put(1.2,0.4){\vector(0,1){0.6}}
\put(2.3,1.4){\footnotesize Optimize ${\bf{A}}_k$}
\put(4.6,0.6){\footnotesize Optimize ${\bf{\lambda}}$}
\put(2.3,0.4){\footnotesize Optimize ${\bf{B}}_k$}
\put(0.2,0.6){\footnotesize Optimize ${\bf{p}}$}
%\put(0.4,0){\footnotesize Figure 1. Block diagram of the joint Tx-Rx
%beamforming scheme.}
\end{picture}
\caption{Block diagram of the joint Tx-Rx beamforming scheme}
\label{Fig 1}
\end{figure}

When ${\bf{p}}$ and ${\bf{B}}_k (k = 1,\ldots,K)$ are fixed,
optimize ${\bf{A}}_k (k = 1,\ldots,K)$ to maximize $SINR_{k,j}^{DL}
(k = 1,\ldots,K,j = 1,\ldots,L)$. Then observing (4), it is a
generalized Rayleigh quotient problem, and its solution is
\begin{equation}
\begin{array}{cll}
 {\bf{a}}_{k,j}  &=& {\bf{\tilde a}}_{k,j} /||{\bf{\tilde a}}_{k,j} ||_2  \\
 {\bf{\tilde a}}_{k,j}  &=& \xi _{\max } ({\bf{R}}_{k,j}^{s,DL} ,{\bf{R}}_{k,j}^{I + n,DL} ) \\
\end{array}
\end{equation}

where $\xi _{\max}({\bf{X}}$,${\bf{Y}})$ is the dominant generalized
eigenvector of the matrix pair $({\bf{X}}$,${\bf{Y}})$. When
${\mbox{\boldmath{$\lambda$}}}$ and ${\bf{A}}_k (k = 1,\ldots,K)$
are fixed, in the same way, ${\bf{B}}_k (k = 1,\ldots,K)$ can be
obtained by
 \begin{equation}
\begin{array}{cll}
 {\bf{b}}_{k,j}  &=& {\bf{\tilde b}}_{k,j} /||{\bf{\tilde b}}_{k,j} ||_2  \\
 {\bf{\tilde b}}_{k,j}  &=& \xi _{\max } ({\bf{R}}_{k,j}^{s,UL} ,{\bf{R}}_{k,j}^{I + n,UL} ) \\
 \end{array}
\end{equation}

The proposed algorithm is summarized in the following.

\begin{algorithm}
\begin{algorithmic}

\STATE{\tt\small Initialize ${\bf{B}}_k^{(0)} (k = 1,\ldots,K)$ and
${\bf{p}}^{(0)}$ randomly. \vspace{3pt} \\ Set the noise vector of
the virtual uplink to ${\bf{w}}$. \vspace{3pt}\\ $n=0$}\vspace{3pt}

\STATE{\tt\textbf{1)}{\tt\small\parindent 6mm \bf{Update in the
primal downlink}. }\vspace{3pt}

\STATE{\tt
a) {\tt\small{\parindent 6mm  Calculate ${\bf{A}}_k^{(n +
1)} (k = 1,\ldots,K)$ from \vspace{1pt}

 ${\bf{B}}_k^{(n)} (k = 1,\ldots,K)$ and ${\bf{p}}^{(n)}$
 using (4)(20).}\vspace{3pt}

b) {\parindent 6mm Calculate ${\bf{C}}^{(n)}$ from
${\bf{B}}_k^{(n)}(k = 1,\ldots,K)$\vspace{1pt}

and ${\bf{A}}_k^{(n + 1)}(k = 1,\ldots,K)$ using
(5)(8)(10).}\vspace{3pt}

c) {\parindent 6mm Solve ${\mbox{\boldmath{$\lambda$}}}^{(n)}  =  -
(({\bf{C}}^{(n)} )^T )^{ - 1} {\bf{w}}$} \vspace{3pt} }}}

\STATE{\tt\textbf{2)}{\tt\small\parindent 6mm \bf{Update in the
virtual uplink}. }\vspace{3pt}

\STATE{\tt a) {\tt\small{\parindent 6mm Calculate ${\bf{B}}_k^{(n +
1)} (k = 1,\ldots,K)$ from \vspace{1pt}

 ${\bf{A}}_k^{(n + 1)} (k = 1,\ldots,K)$ and ${\mbox{\boldmath{$\lambda$}}}^{(n)}$
 using (16)(21).}\vspace{3pt}

b) {\parindent 6mm Calculate ${\bf{C}}^{(n + 1)}$
  from ${\bf{B}}_k^{(n + 1)} (k = 1,\ldots,K)$ \vspace{1pt}

and ${\bf{A}}_k^{(n + 1)} (k = 1,\ldots,K)$ using
(5)(8)(10).}\vspace{3pt}

c) {\parindent 6mm Solve ${\bf{p}}^{(n + 1)}  =  - ({\bf{C}}^{(n +
1)} )^{ - 1} {\bf{d}}$ \vspace{1pt}

$n = n+1$} \vspace{3pt} }}}

\STATE{\tt\textbf{3)}{\tt\small\parindent 6mm \bf{Repeat 1) and 2)
until \vspace{1pt}}

$\sum\limits_{k = 1}^K {||{\bf{A}}_k^{(n)}  - {\bf{A}}_k^{(n + 1)}
||_F }  + \sum\limits_{k = 1}^K {||{\bf{B}}_k^{(n)}  -
{\bf{B}}_k^{(n + 1)} ||_F }  \le \varepsilon $ }}.\vspace{1pt}

\STATE{}{\tt \small \parindent 6mm

\hspace {5mm} In the simulation, we set $\varepsilon  = 0.0001$.}

\end{algorithmic}
\end{algorithm}

By iteration,
${({\bf{A}}_k^{(n+1)},{\bf{B}}_k^{(n+1)},{\bf{p}}^{(n+1)})}{(k=1,\ldots,K)}$
converges to the optimal solution to the optimization problem (2).

Once any element in ${\bf{p}}^{(n + 1)}$ is negative, which
indicates the post-SINR goals
${\gamma}_{k,j}(k=1,\ldots,K,j=1,\ldots,L)$ can not be attained,
${\gamma}_{k,j}$ should be decreased to relax the post-SINR
constraints. When $\bf{w}=\bf{1}$, the proposed algorithm is similar
with the one in \cite{11}.

\begin{figure}
\centering
\includegraphics[width=3in]{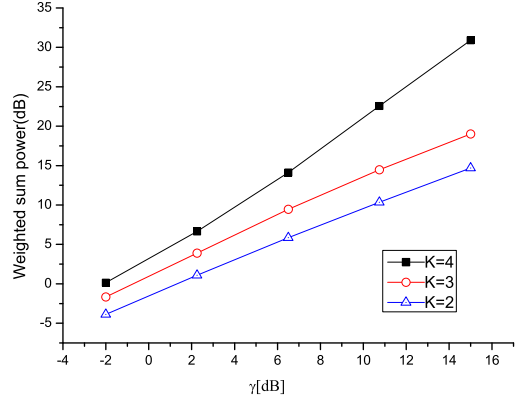}
\caption{Total transmit power versus SINR goal $\gamma$, when
$K=2$,$3$,$4$} \label{Fig 2}
\end{figure}

\begin{figure}
\centering
\includegraphics[width=3in]{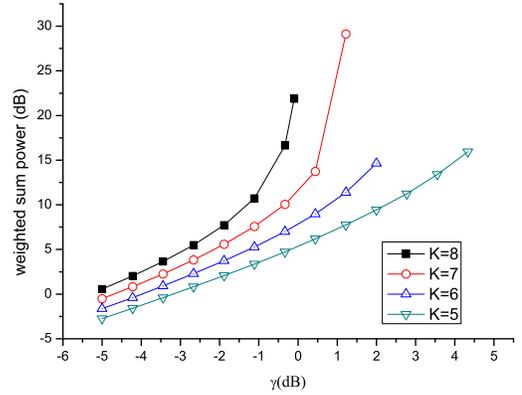}
\caption{Total transmit power versus SINR goal $\gamma$, when
$K=5$,$6$,$7$,$8$.} \label{Fig 3}
\end{figure}

\section{Simulation result}
In this section, we assume that a BS with $8$ antennas ($M=8$) is
communicating with $K$ MS's each with $2$ antennas,
$(N_1=\ldots=N_K=2)$. Also we assume that the number of substreams
of each MS is the same, equal to $2$,$(L_1=\ldots=L_K=2)$. QPSK is
employed in the simulation and no forward error coding is
considered. The post-SINR goals for all substreams are $\gamma
({\gamma_{k,j}=\gamma})$. Additionally we assume MS$_1$ is an edge
user in a cell, according to the previous section, a higher weight
should be assigned to it to mitigate the adjacent cell interference.
Thus, the weight vector is set to ${\bf{w}} = [w,w,1,\ldots,1]^T$,
where $w$ is the weight corresponding to the two substreams of
MS$_1$ and $w>1$.

\begin{figure}
\centering
\includegraphics[width=3in]{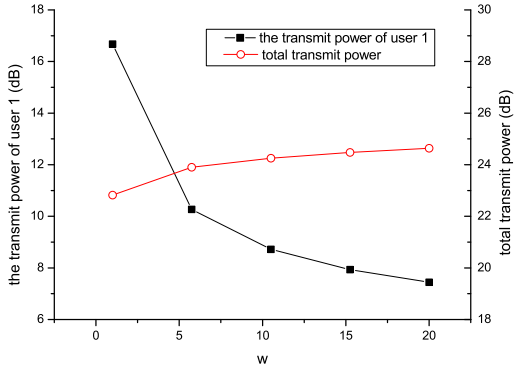}
\caption{When $K=4$, $\gamma=10$dB, the transmit power of MS$_1$ and
total power versus the weight $w$.} \label{Fig 4}
\end{figure}

Fig. \ref{Fig 2},\ref{Fig 3} plot the curves of the total transmit
power ${\sum\limits_{k,j}{p_{k,j}}}$ versus the post-SINR goal
$\gamma$, when $w=5$. In Fig. \ref{Fig 2}, $K=2,3,4$, the system
configuration satisfies $M \ge KL$, the multiuser interference,
thus, can be effectively suppressed through the beamforming
\cite{4}. Under the circumstance, increasing the transmit power of
any user has nearly no effect to the post-SINR of other users.
Therefore, all the substreams can attain relatively high post-SINR.
In Fig. \ref{Fig 3}, $K = 5,6,7,8$, $M \ge KL$ does not hold any
more. Consequently, the multiuser interference can not be
effectively mitigated, which means any enhancement in the transmit
power of any user is very likely to deteriorate the post-SINR of
other users. As shown in Fig. \ref{Fig 3}, with the user number
increasing, the available post-SINR of each user is decreased. When
$K=8$, only $0$ dB post-SINR can be attained. In these two figures,
the total transmit power increases with the number of users and the
post-SINR goal $\gamma$. Especially when $K=7,8$ and $\gamma \geq 0$
dB in Fig. \ref{Fig 3}, due to the residual multiuser interference,
the slopes of the curves are much steeper than that in Fig. \ref{Fig
2} where the multiuser interference is negligible. And the steeper
the curves are, the more power would be paid for the unit increase
of the post-SINR of each user.

Fig. \ref{Fig 4} shows the curves of the transmit power
${\sum\limits_j{p_{1,j}}}$ of MS$_1$ and the total transmit power
${\sum\limits_{k,j}{p_{k,j}}}$ versus the weight $w$, when $K=4$ and
$\gamma = 10$dB. The left vertical axis is corresponding to the
transmit power of MS$_1$ and the right one is to the total transmit
power. Obviously, as the $w$ is increasing, the transmit power of
MS$_1$ is decreasing while the total power is increasing, because
the optimization object is to minimize the weighted sum power
${\sum\limits_{k,j}{w_{k,j}p_{k,j}}}$. Moreover, when $w$ changing
from $1$ to $20$, the transmit power of MS$_1$ decreases almost
$10$dB, however the total power increases only about $1$ dB, which
demonstrates that the proposed algorithm adapts the power allocation
policy very effectively with negligible penalty on performance.

\section{conclusion}
In this paper, we investigate the joint Tx-Rx design for the
downlink multiuser spatial multiplexing MIMO system. We show, first,
the uplink-downlink duality has the following characteristics: 1) In
both of the primal downlink and the virtual uplink, the substreams
can attain the same post-SINR goal; 2) The beamforming matrices are
common in both of the primal downlink and the virtual uplink.

Based on the duality, a joint Tx-Rx beamforming scheme is proposed.
Simulation results demonstrate that the scheme can not only satisfy
the post-SINR constraints which guarantee the performance of the
communication links, but also easily adjust the power distribution
among users by changing the weights correspondingly, which can be
used to diminish the power of the edge users in a cell to alleviate
the adjacent cell interference.


\begin{thebibliography}{1}
\bibitem{1}
I. Telatar, "Capacity of multi-antenna Gaussian channels",
\emph{Eur. Trans. Telecommun}, vol. 10, no. 6, pp.585-595, Nov./Dec.
1999.
\bibitem{2}
Q. Caire and S. Shamai, "On the achievable throughput of a
multiantenna gaussian broadcast channel", \emph{IEEE Trans. Inform.
theory}, vol. 49, no. 7, pp.1691-1706, July 2003.
\bibitem{3}
Q. Spencer, A. Swindlehurst and M. Haardt, "Zero-forcing methods for
downlink spatial multiplexing in multiuser mimo channels",
\emph{IEEE Trans. Signal Processing}, vol. 52, no. 2, pp.461- 471,
Feb. 2004.
\bibitem{4}
Z.G. Pan,K.K. Wong and T.S. Ng, "Generalized multiuser orthogonal
space division multiplexing", \emph{IEEE Trans. Wireless Commun.},
vol. 3, no. 6, pp.1969- 1973, Nov. 2004.
\bibitem{5}
J. Zhang, Y. Wu, S. Zhou and J. Wang, "Joint linear transmitter and
receiver design for the downlink of multiuser MIMO systems",
\emph{IEEE commun. Lett}, vol.9, pp.991-993, Nov. 2005.
\bibitem{6}
F. Rashid-Farrokhi, L. Tassiulas, and K.J Liu, "Joint optimal power
control and beamforming in wireless networks using antenna array",
\emph{IEEE Trans. Commun.}, vol. 46, no. 11, pp.1313-1324, Nov.
1998.
\bibitem{7}
F. Rashid-Farrokhi F., K.J. Liu and L. Tassiulas, "Transmit
beamforming and power control for cellular wireless systems",
\emph{IEEE J. Sel. Areas Commun.}, vol. 16, no. 8, pp.1437-1450,
Oct. 1998.
\bibitem{8}
J.H. Chang, L. Tassiulas and F. Rashid-Farrokhi, "Joint transmitter
receiver diversity for efficient space division multiaccess",
\emph{IEEE Trans. Wireless Commun.}, vol. 1, no. 1, pp.16-27, Jan.
2002.
\bibitem{9}
S. Boyd and L. Vandenberghe, Convex Optimization, \emph{Cambridge:
U.K.} Cambridge University Press, 2004.
\bibitem{10}
B. Song, R.L. Cruz and B.D. Rao, "Network Duality for Multiuser MIMO
Beamforming Networks and Applications", \emph{IEEE Trans. Commun.},
vol.55, no.3, pp.618-629, Mar. 2007.
\bibitem{11}
A.M. Khachan, A.J. Tenenbaum and R.S. Adve, "Linear Processing for
the Downlink in Multiuser MIMO Systems with Multiple Data Streams",
\emph{IEEE ICC'06}, vol. 9, pp.4113-4118, June 2006.
\end{thebibliography}
\end{document}